%Paper: hep-th/9504104
%From: Ali Yegulalp <yegulalp@puhep1.Princeton.EDU>
%Date: Thu, 20 Apr 1995 11:58:53 -0400 (EDT)

\input harvmac.tex      %% input your version
\font\blackboard=msbm10 \font\blackboards=msbm7
\font\blackboardss=msbm5
\newfam\black
\textfont\black=\blackboard
\scriptfont\black=\blackboards
\scriptscriptfont\black=\blackboardss
\def\blackb#1{{\fam\black\relax#1}}
\def\BR{{\blackb R}}
\def\BZ{{\blackb Z}}

\def\cR{{\cal R}}
\def\cW{{\cal W}}
\def\vol{\mathop{\rm vol}}
\def\sgn{\rm sgn}
\def\up{\mathop{\rm up}}
\def\Tr{\mathop{\rm Tr}}

\def\tw{\tilde{w}}
\def\tV{\tilde{V}}
\def\tX{\tilde{X}}
\def\tepsilon{\tilde{\epsilon}}

\def\bz{\bar{z}}
\def\bw{\bar{w}}
\def\brho{\bar{\rho}}
\def\bX{\bar{X}}
\def\bV{\bar{V}}
\def\bpar{\bar{\partial}}
\def\balpha{\bar{\alpha}}
\def\bp{\bar{p}}
\def\bq{\bar{q}}
\def\bL{\bar{L}}
\def\pD{D^{\prime}}
\def\pphi{\phi^{\prime}}
\def\hg{\hat{g}}
\def\>{\rangle}
\def\<{\langle}

%%%%%%%% references %%%%%%%%
%\def\Pol{J. Polchinski, Nucl. Phys. {\bf B242} (1984) 345.}
\lref\acny{A. Abouelsaood, C. G. Callan, C. R. Nappi
  and S. A. Yost, Nucl. Phys. B280 [FS18] (1987) 599.}
\lref\clny{C. G. Callan, C. Lovelace, C. R. Nappi
  and S. A. Yost, Nucl. Phys. B 293 (1987) 83;
  Nucl. Phys. B 308 (1988) 221.}
\lref\cardy{John L. Cardy, Nucl. Phys. B324 (1989) 581.}
\lref\ali{Ali Yegulalp, Phys. Lett. B 328 (1994) 379.}
\lref\ckmy{C. G. Callan, I. R. Klebanov, J. M. Maldacena and
   A. Yegulalp, PUPT-1528 (1995); hep-th/9503014.}
\lref\cklm{Curtis G. Callan, Igor R. Klebanov, Andreas W. W. Ludwig
   and Juan M. Maldacena,
   Nucl. Phys. B 422 (1994) 417.}
\lref\freed{Denise E. Freed, MIT-CTP-2422 (1995);
  hep-th/9503065.}
\lref\cff{Curtis G. Callan, Andrew G. Felce and Denise E. Freed,
  Nucl. Phys. B 392 (1993) 551.}
\lref\go{Peter Goddard and David Olive,
   Int. J. Mod. Phys. A 1 (1986) 303.}
\lref\cornwell{J. F. Cornwell, Group theory in physics
  (Academic Press, San Diego, 1990).}
\lref\var{ V. S. Varadarajan, Lie groups, Lie algebras,
  and their representations (Springer-Verlag, New York, 1984).}
\lref\fls{P. Fendley, A. Ludwig and H. Saleur,
  cond-mat/9408068.}
\lref\cl{A. O. Caldeira and A. J. Leggett, Physica 121 A (1983)
  587;
  Phys. Rev. Lett. 46 (1981) 211;
  Ann. of Phys. 149 (1983) 374.}
\lref\ct{C. G. Callan and L. Thorlacius, Nucl. Phys. B 329
  (1990) 117.}
\lref\al{I. Affleck and A. Ludwig, Nucl. Phys. B 352
  (1991) 841.}
\lref\cd{C. G. Callan and S. Das, Phs. Rev. Lett. 51
  (1983) 1155.}
\lref\as{Ian Affleck and Jacob Sagi,
  Nucl. Phys. B 417 (1994) 374.}
\lref\hofstadter{M. Ya Azbel', Zh. Eksp. Teor. Fiz. 46 (1964)
  929;  D. R. Hofstadter, Phys. Rev. B 14 (1976) 2239;
  G. H. Wannier, Phys. Status Solidi B 88 (1978) 757.}

%%%%%%%%%%%%%%%%%%%%%%%%%%%%

\Title{PUPT - 1538, April 1995}
{\vbox{\centerline{New Boundary Conformal Field Theories}
\centerline{Indexed by the Simply-Laced Lie Algebras} }}

\bigskip
\bigskip

\centerline{Ali Yegulalp\footnote{*}{
  yegulalp@puhep1.princeton.edu}}
\medskip\centerline{\it Department of Physics, Princeton University}
\centerline{\it Princeton, NJ 08544}
\bigskip\bigskip

\vskip .25in

\centerline{\bf Abstract}

We consider the field theory of $N$ massless bosons which are free
except for an interaction localized on the boundary of their 1+1
dimensional world.  The boundary action is the sum of two pieces:
a periodic potential and a coupling to a uniform abelian gauge
field.
Such models arise in open string theory and dissipative
quantum mechanics, and possibly in edge state tunneling in the
fractional quantized Hall effect.
We explicitly show that conformal invariance is unbroken for certain
special choices of the gauge field and the periodic potential.
These special cases are naturally indexed by semi-simple, simply
laced Lie algebras.  For each such algebra, we have a discrete
series of conformally invariant theories where the potential
and gauge field are conveniently given in terms of
the weight lattice of the algebra.
We compute the exact boundary state for these theories, which
explicitly shows the group structure.  The partition function
and correlation functions are easily computed using the boundary
state result.

\Date{}

\eject
%\noblackbox

\newsec{Introduction}
Free field theory on a 1+1 dimensional manifold with boundary has been
the focus of a number of recent works.
With the proper choice of boundary
interactions, such theories can describe dissipative quantum mechanics
\refs{\cl,\ct},
tunneling between quantum Hall edges \fls{},
impurity scattering \al{},
open strings in background fields \refs{\clny,\acny}
, and monopole catalysis \refs{\cd,\ali,\as}.
Typically, the interesting physics lies mostly in the infrared limit
of these theories, and so it is of considerable interest to identify
the conformally-invariant fixed points to which they flow.

We will study the following Euclidean action:
\eqna\action
$$\eqalignno{
S &= S_{bulk} + S_{gauge} + S_{pot}, &\action{a}\cr
S_{bulk} &= {1\over{8\pi}} \int_0^T d\tau \int_0^{l}
     d\sigma \Bigl(\partial_{\tau}X \cdot \partial_{\tau}X +
              \partial_{\sigma}X \cdot \partial_{\sigma}X \Bigr),
    &\action{b} \cr
S_{gauge} &= {i\over{8\pi}}\int_0^T d\tau \, X(0,\tau) \cdot B
       \cdot \partial_{\tau} X(0,\tau), &\action{c}\cr
S_{pot} &= \int_0^T d\tau \,V(X(0,\tau)),&\action{d}
 \cr}
$$
where
$$\eqalign{
V(X) &= \sum_k \Bigl(V_{\omega_k} e^{i \omega_k\cdot X} +
              V_{-\omega_k} e^{-i \omega_k \cdot X} \Bigr).
}
$$
$X(\sigma,\tau)$ is a real-valued $N$ component bosonic field,
$B$ is a real anti-symmetric matrix, and the $\omega_k$ are
real $N$ component vectors.  The complex parameters $V_{\omega}$
obey the relation $V_{-\omega} = \bV_{\omega}$, so that $V(X)$
is a real potential.  $S_{bulk}$ describes a free massless
theory on the interval $0 \leq \sigma \leq l$.
The terms $S_{gauge}$ and $S_{pot}$ are perturbations localized
at the $\sigma=0$ boundary.  In principle, we could also add similar
perturbations localized at $\sigma=l$, but we will omit such terms
since our later analysis shows
that we can treat each boundary independently.

 From the point of view of string
theory, this action describes a bosonic open string whose endpoints
feel the presence of a background gauge field and periodic potential.
The conformal fixed
points of the action correspond to background field configurations
which solve the classical open string field equations.

Our action can also be used to study the dissipative quantum mechanics
(DQM) of a non-relativistic particle subject to the periodic potential
$V(X)$ and magnetic field $B$ (the Wannier-Azbel-Hofstadter
model \hofstadter{}).  The correspondence between open string theory and DQM
has been outlined in detail \refs{\ct,\cff},
and may be summarized as follows:
the endpoint of the string corresponds to the location of the DQM particle.
The fields felt by the string endpoint correspond to the fields felt by the
particle.
As the string endpoint moves, it loses energy by exciting
the modes of the string, which corresponds to the DQM particle losing
energy by exciting a dissipative bath of oscillators.
The mapping between the open string and DQM becomes exact in the
infrared limit, and so it is of greatest interest to study the
conformal fixed points of the action:  at these points,
the correlators of $X$ show scaling behaviour,
indicating that we are at a transition point between localized
and delocalized behavior for the DQM particle.

Our action $S$ subsumes and generalizes some special cases considered
in a number different papers.
The oldest preceding paper \acny{} studies the
quadratic action $S_{bulk}+S_{gauge}$, i.e., the $V(X)=0$
limit of our model.  The exact solution of the $V=0$ case shows
that the action is conformally invariant for generic values of $B$.
In \cklm{}, the authors considered the simplest $V \neq 0$ case:
$N=1$, $B=0$, and $V(X)=V_0\cos({X/\sqrt{2}})$.
That action proves to be conformally invariant for
any value of $V_0$, thanks to the presence of an $SU(2)_1$
Kac-Moody symmetry.  A number of papers \refs{\cff,\ckmy,\freed}
have been devoted
to the next simplest case, with $N=2$, $B_{ij}=b\epsilon_{ij}$,
and $V(X)=V_0(\cos(aX)+\cos(aY))$.  In this case, the analysis
depends in a critical way on the parameter $b$.  If $b$ is an integer,
the theory is conformally invariant for any value of $V_0$,
provided we set $a=\sqrt{1+b^2\over 2}$.  In direct
analogy with the $N=1$ case, an $SU(2)_1 \times SU(2)_1$ symmetry
appears in the solution.  When $b$ is not an integer, things are not
so simple: if we set $a=\sqrt{1+b^2\over 2}$, then $V_0=0$ is the
only fixed point.  If we take $a<\sqrt{1+b^2\over 2}$, then
perturbative calculations indicate that $V_0=0$ becomes unstable,
and $V_0$ flows to a finite fixed point value in the infrared.
No Kac-Moody algebra arises for non-integer values of $b$, and
consequently no exact results are available.

In this paper, we seek to generalize the results obtained for
integer $b$ in the $N=2$ model.
Our approach is to choose fixed values for $B$ and
$\omega_k$ and turn on $V(X)$ as a perturbation.
Using boundary state technology, we will show that there
is a series of special values for $B$ and $\omega_k$ such
that the perturbation $V(X)$ is exactly marginal.
In these special cases, one may exactly solve
the theory due to the presence of a level 1, simply laced
affine Kac-Moody algebra.
Furthermore, for any semi-simple, simply laced Lie algebra $g$,
there exists a discrete series of choices for $B$ and $\omega_k$
that guarantee conformal invariance.  The constants $V_{\omega_k}$
parameterize a manifold of fixed points isomorphic to the corresponding
Lie group $G$.
Interestingly,
$B=0$ is allowed only if $g$ is a direct sum of $su(2)$
algebras;  the gauge field interaction is indispensable for realizing
all the other algebras.  We will compute the exact boundary
state for these theories, from which we may easily compute the
partition function and correlation functions.

\newsec{Basic Setup}
Our primary goal is to compute the functional integral
on a cylinder of length $l$ and circumference $T$.  As is well
known \cardy{}, there are two equivalent pictures for the path
integral.  In the first picture, we are computing the partition
function $Z = tr(e^{-TH_{open}})$ for an open string of length $l$
at temperature T;  the Hamiltonian $H_{open}$ is that of an open
string with boundary action described by $S_{gauge}+S_{pot}$.  In the dual
picture, we are computing the amplitude
$Z= \<N|e^{-lH_{closed}}|V\>$, which describes a free closed
string of length $T$ propagating for time $l$ between two
`boundary states'.  The boundary states correspond to the
ends of the cylinder, and they conveniently summarize the
dynamics present at the boundaries of the open string.
In this case, $|N\>$ represents the free Neumann boundary
condition and $|V\>$ represents the boundary condition
induced by $S_{gauge}+S_{pot}$.  Conformal invariance of
the open string action translates into the Virasoro
constraint $(L_n-\bL_{-n})|V\>=0$.

In this paper, we will work in the closed string channel and
focus on computing the boundary state $|V\>$.  Consequently, we
can use the standard closed string chiral mode expansions:
$$\eqalign{
X(z,\bz) = X(z) + \bX(\bz),
}
$$
\eqna\xfield
$$\eqalignno{
X(z) &= q - ip\ln(z) + \sum_{n \neq 0}{i\over n} \alpha_n z^{-n},
\cr
\bX(\bz) &= \bq -i\bp\ln(\bz) + \sum_{n \neq 0}{i\over n}
    \balpha_n \bz^{-n}, \cr
}
$$
where
$$\displaylines{
[q_j,p_k] = [\bq_j,\bp_k] = \qquad i\delta_{j,k}, \cr
[q_j,q_k] = [p_j,p_k]=[\bq_j,\bq_k]=[\bp_j,\bp_k]=[q_j,\bp_k]=
  [\bq_j,p_k] = 0,\cr
[\alpha_{nj},\alpha_{mk}] =
[\balpha_{nj},\balpha_{mk}] = \qquad n\delta_{j,k}\delta_{n+m,0},\cr
[\alpha_{nj},\balpha_{mk}] = 0,\cr
\alpha^{\dagger}_{nj} = \alpha_{-nj},\qquad
 \alpha_{0i} =p_i\cr
z = e^{2\pi (\tau + i\sigma)/T} \qquad \hbox{and} \qquad
\bz = e^{2\pi(\tau - i\sigma)/T}.\cr
}
$$
The Hamiltonian is $H_{closed} = {2\pi\over T}(L_0+\bL_0)$, where
we have the standard Virasoro algebra
$$\eqalignno{
L_0 &= {1\over 2}p^2 + \sum_{n=1}^{\infty}\alpha^{\dagger}_n \cdot
    \alpha_n - {N\over 24},\cr
L_k &= {1\over 2}\sum_{n=-\infty}^{\infty}:\alpha_n
    \cdot \alpha_{k-n}:,\cr
[L_n,L_m] &= (n-m) L_{n+m} + {N\over 12}(n^3-n)\delta_{n+m,0},\cr
}$$
with corresponding expressions for $\bL_n$.

Since $[X(z),\bX(\bw)]=0$, we can generate the closed string Hilbert
space by taking tensor products of states built from left-moving
and right-moving modes.  There is one caveat:  we must require
$X(z)+\bX(\bz)$ to be single-valued as $\sigma\rightarrow
\sigma+T$.  If $X$ is non-compact, then only states with $p=\bp$
are allowed.

We can also take $X(z,\bz)$ to be
compactified on a lattice $\Lambda$, corresponding to the identification
$X+\bX\equiv X+\bX + \lambda$, where $\lambda\in\Lambda$.
Our Hilbert space is then subject to the restriction
$p-\bp \in {1\over 2\pi}\Lambda$.  Note that we regard this as a
restriction on the Hilbert space, not on the operators $p$ and $\bp$.
Compactifying $X$ also discretizes the momentum spectrum,
giving us the additional restriction $p+\bp\in 4\pi\Lambda^{*}$,
where $\Lambda^{*}$ is the dual of $\Lambda$.  A summary of lattice
facts and conventions is given in the appendix.

We will be interested in both the compact and non-compact cases.
To avoid writing two copies of all future results, we make
the simple observation that the non-compact case is actually
subsumed in the compact case if we allow the degenerate lattice
$\Lambda=\{0\}$;  the dual lattice in that case is $\Lambda^*=
\BR^{N}$, so the momentum spectrum is indeed continuous.

\newsec{Neumann, Dirichlet, and Gauge Field Boundary States}
Before tackling the computation of the boundary state $|V\>$,
it will be helpful to consider some simpler cases:  Neumann,
Dirichlet, and gauge field boundary states.  Once we have the
gauge field state $|B\>$, we can express $|V\>$ as a perturbation
of $|B\>$.

We will begin by considering just the action $S_{bulk}$,
which leads to Neumann boundary conditions on $X$ in the
open string channel. In the closed string channel, this means
that the boundary state must satisfy
$\partial_{\tau}X(\sigma,\tau)|N\>
= 0$ at $\tau=0$.  In terms of oscillator modes, we have
\eqna\Ncond
$$\eqalignno{
(p+\bp)|N\> &= (\alpha_n+\balpha_{-n})|N\>=0.&\Ncond{} \cr
}$$
This condition yields the state
\eqna\Nstate
$$\eqalignno{
|N\> &= 2^{-N/4} \sqrt{\vol{(\Gamma)}}\;
   e^{-\sum_{n=1}^{\infty}{1\over n}\balpha^{\dagger}_n
     \cdot \alpha^{\dagger}_n}
   \sum_{\gamma\in\Gamma} |p={\gamma\over\sqrt{2}},\;\;
   \bp ={-\gamma\over\sqrt{2}}\>, &\Nstate{}\cr
}
$$
where $\Gamma={1\over 2\pi\sqrt{2}}\Lambda$, $\Lambda$
is the lattice of compactification for $X$, and
$|p=k, \bp = \bar{k}\>$ denotes the oscillator ground state
with momenta $p=k$ and $\bp=\bar{k}$.

Strictly speaking, equation \Ncond{} does not determine
the overall normalization of $|N\>$.  It also does not determine
which momenta contribute to the sum, other than restricting
$p+\bp=0$.  One can nevertheless verify that equation \Nstate{}
is correct by directly computing the open string partition
function with Neumann boundary conditions at both ends, then
comparing the result with the partition function computed using
the boundary state $|N\>$.  Explicitly, the direct open string
calculation yields
\eqn\ZNNopen{
Z_{NN} = w^{-N/24} \prod_{n=1}^{\infty}(1-w^n)^{-N}
   \sum_{\gamma^{*}\in\Gamma^{*}} w^{(\gamma^{*})^2},
  \quad\hbox{where}\quad w=e^{-\pi T/l},
}
while the closed string boundary state method yields
\eqna\ZNNclosed
$$\eqalignno{
Z_{NN} &= \<N|e^{-{2\pi l \over T}(L_0+\bL_0)}|N\> &\ZNNclosed{} \cr
       &= 2^{-N/2}\vol{(\Gamma)}\tw^{-N/12}
    \prod_{n=1}^{\infty}(1-\tw^{2n})^{-N}
    \sum_{\gamma\in\Gamma} \tw^{\gamma^2/ 2},
   \quad \hbox{where} \quad \tw = e^{-2\pi l/T}.
}
$$
Poisson resummation establishes that
\ZNNopen{} and \ZNNclosed{} are equivalent.

Similar considerations lead to the Dirichlet boundary
equations
\eqn\Dcond{
(p-\bp)|D\> = (\alpha_n-\balpha_{-n})|D\> = 0
}
and the Dirichlet boundary state
\eqna\Dstate
$$\eqalignno{
|D\> &= 2^{-N/4} \sqrt{\vol{(\Gamma^{*})}}\;
   e^{\sum_{n=1}^{\infty}{1\over n}\balpha^{\dagger}_n
     \cdot \alpha^{\dagger}_n}
   \sum_{\gamma^{*}\in\Gamma^{*}} |p={\gamma^{*}\over\sqrt{2}},\;\;
   \bp ={\gamma^{*}\over\sqrt{2}}\>. &\cr
}
$$
Given equations \Ncond{} and \Dcond{}, one may easily
verify the conformal invariance constraints
$(L_n-\bL_{-n})|N\> = (L_n-\bL_{-n})|D\>=0$.

Finally, we will consider the action $S_{bulk}+S_{gauge}$, which
results in the gauge field boundary state $|B\>$.
In a compactified theory, we must consider what happens
to $S_{gauge}$ when we make the physically unobservable shift
$X \rightarrow X+\lambda$, with $\lambda\in\Lambda$.
Looking at equation \action{c}, we see that
$S_{gauge} \rightarrow S_{gauge} + {i\over 8\pi}\lambda^{\prime} \cdot
B \cdot \lambda$, where $\lambda,\lambda^{\prime}\in\Lambda$.
The path integral will be invariant under such shifts,
provided that
\eqn\Bsense{
{1\over 2} B \Gamma \subset \Gamma^{*}.
}
Varying the action $S_{bulk}+S_{gauge}$ gives a linear constraint
on $X$ at the boundary, which translates into the closed
string condition $((1+B)\cdot\alpha_n +
(-1+B)\cdot\balpha^{\dagger}_n)|B\>=0$.  This leads to the
solution
$$\eqalignno{
|B\> =& \;\;2^{-N/4} \sqrt{\det(1+B)\vol{(\Gamma)}} \cr
  &\;\;e^{-\sum_{n=1}^{\infty} {1\over n} \balpha^{\dagger}_n
    \cdot M \cdot \alpha^{\dagger}_n}
  \sum_{\gamma\in\Gamma} | p={1\over\sqrt{2}}(1-B)\gamma,
    \;\; \bp=-{1\over\sqrt{2}}(1+B)\gamma\>,
}
$$
where we define the orthogonal matrix
$$M = {1+B\over 1-B}.$$

In the next section, we will compute $|V\>$ as a perturbation
on $|B\>$.  To this end, it is useful to rearrange our expression
of $|B\>$ into the following form:
\eqn\Bdef{
|B\> = P_{\sqrt{2}\Gamma}(p-\bp)S_BR|\pD\>.
}
We have introduced several new objects;  first we have the
projection operator
\eqn\projdef{
P_{\Omega}(x) =\cases{1,&if $x \in\Omega$;\cr
   0,&otherwise.\cr}
}
Next, we define the chiral rotation and reflection operators
$S_B$ and $R$:
$$\displaylines{
S_BX(z)S^{\dagger}_B = M\cdot X(z), \qquad
S_B\bX(\bz)S^{\dagger}_B = \bX (\bz ),\cr
RX(z)R^{\dagger} = -X(z), \qquad
R\bX(\bz)R^{\dagger} = \bX (\bz), \cr
S_B^{\dagger} S_B =1\quad\hbox{and} \quad R=R^{\dagger}=R^{-1}.
}
$$
Finally, we define a boundary state which is almost the same as
the Dirichlet boundary state:
\eqn\Dprimedef{
|\pD\> = 2^{-N/4} \sqrt{\det(1+B)\vol{(\Gamma)}}
   e^{\sum_{n=1}^{\infty} {1\over n} \balpha^{\dagger}_n
       \cdot \alpha^{\dagger}_n}
   \sum_{\mu\in\Upsilon} |p=\mu,\;\;\bp=\mu\>.
}
The set $\Upsilon$ can be any lattice that satisfies
${1\over\sqrt{2}}(1+B)\Gamma\subset\Upsilon$;  the projection
operator $P$ removes any extra states.  The freedom to chose
$\Upsilon$ will be useful when we compute $|V\>$.

Conformal invariance of $|B\>$ is readily verified.
 From the definitions above,
it is easy to see that $L_n$ and $\bL_n$ commute with
the operators $S_B$,$R$, and $P$, so we have
\eqn\Bconf{
(L_n-\bL_n)|B\> = P_{\sqrt{2}\Gamma}(p-\bp)S_BR
  (L_n-\bL_n)|\pD\> = 0.}

\newsec{The Boundary State with Gauge Field and Potential}
We will now compute the boundary state $|V\>$ for our complete
action $S=S_{bulk}+S_{gauge}+S_{pot}$.  The basic idea is to
regard $S_{bulk}+S_{gauge}$ as the free theory with $S_{pot}$
as a perturbation.  For $S_{pot}$ to be well-defined in a
compactified theory, we must require
\eqn\omsense{
\omega_k \in {1\over\sqrt{2}}\Gamma^{*}.
}
Assuming this is true, we have
\eqna\pertdef
$$\eqalignno{
|V\> &= e^{-\int_{0}^{T}V(X(\tau=0,\sigma)) d\sigma} |B\> \cr
    &= e^{{iT\over 2\pi}\oint {dz\over z} V(X(z)+\bX({1\over z}))}
      \;\;P_{\sqrt{2}\Gamma}(p-\bp)S_BR|\pD\>.
}
$$
The contour of integration is the circle $|z|=1$.  Since the
projection operator only depends on $p-\bp$, it commutes with
the potential $V(X(z)+\bX({1\over z}))$.  Shifting the operators
around, we get
\eqn\pert{
|V\> =
P_{\sqrt{2}\Gamma}(p-\bp)S_BR \;
e^{{iT\over 2\pi}\oint {dz\over z} V(-M^{t}\cdot X(z) +
   \bX({1\over z}))} |\pD\>.
}
Expanding the exponential, we obtain
\eqna\pertii
$$\eqalignno{
|V\> =&
P_{\sqrt{2}\Gamma}(p-\bp)S_BR \;
\sum_{n=0}^{\infty} {1\over n!} \Bigl({iT\over 2\pi}\Bigr)^{n}
  \oint {dz_1\over z_1} \cdots \oint {dz_n\over z_n}\;\;\;
  \sum_{\nu_j \in \{\pm\omega_k \} } V_{\nu_1} \cdots V_{\nu_n} \cr
  &e^{i\nu_1\cdot (-M^{t}\cdot X(z_1) + \bX({1\over z_1}))}
  \cdots
  e^{i\nu_n\cdot (-M^{t}\cdot X(z_n) + \bX({1\over z_n}))}
|\pD\>.&\pertii{}\cr
}
$$

\subsec{Evaluation of a Typical Term}
Let us focus our attention on the exponential factors in \pertii{}.
A typical term is of the form
\eqn\typ{
|typ\> =
e^{i\nu_1\cdot (-M^{t}\cdot X(z_1) + \bX({1\over z_1}))}
  \cdots
  e^{i\nu_n\cdot (-M^{t}\cdot X(z_n) + \bX({1\over z_n}))}
|\pD\>,
}
where $\nu_1,\cdots \nu_n \in \{\pm\omega_k\}$.
With some manipulation, we can put \typ{} in a form that
will allow us to evaluate \pertii{} very explicitly.
All the identities that we will apply to \typ{} are easily
derived from the basic Campbell-Hausdorff identities
\eqn\CampHaus{
e^A e^B = e^B e^A e^{[A,B]} \quad\hbox{and}\quad
e^{A+B} = e^A e^B e^{-{1\over 2}[A,B]},
}
which are true whenever $[A,[A,B]] = [B,[A,B]]=0$.

We begin by separating the left movers from the right movers:
\eqn\typii{
|typ\> = e^{-i\nu_1\cdot M^t \cdot X(z_1)} \cdots
         e^{-i\nu_n\cdot M^t \cdot X(z_n)} \;\;
         e^{i\nu_1\cdot \bX({1\over z_1})} \cdots
         e^{i\nu_n\cdot \bX({1\over z_n})}
    |\pD\>.
}
Next, we note that $|\pD\>$ converts right-movers into
left-movers in the following way:
\eqn\lrid{
e^{i\nu\cdot\bX({1\over z})} |\pD\> =
e^{-i\nu\cdot X(z)} |\pD\>.
}
This identity follows from \Dcond{} and \CampHaus{},
providing that the lattice $\Upsilon$ includes the momentum
$\nu$.  Since we are free to make $\Upsilon$ as dense as we
like, we assume it contains $\nu$.

Applying \lrid{} to \typii{}, we see that we can convert
the rightmost
right-moving exponential into a left-moving
one, then commute it to the left of all the remaining
right-movers.  Continuing until we eliminate all the
right-movers, we obtain
\eqn\typiii{
|typ\> = e^{-i\nu_1\cdot M^t \cdot X(z_1)} \cdots
         e^{-i\nu_n\cdot M^t \cdot X(z_n)}
         e^{-i\nu_n\cdot X(z_n)} \cdots
         e^{-i\nu_1\cdot X(z_1)}
|\pD\>.
}

Now we introduce a normal ordering convention which puts
$q$ to the left of $p$ and $\alpha^{\dagger}_n$ to the
left of $\alpha_n$ for $n>0$.  Applying \CampHaus{}, we see
that
\eqn\normord{
e^{i\nu\cdot X(z)} = (\epsilon z)^{{1\over 2} \nu^2}
  :e^{i\nu\cdot X(z)}:,
}
where $\epsilon$ is a short distance cutoff introduced to regulate
the divergences due to contractions of $X(z)$ with itself.
To be precise, we have regulated the short distance behaviour
by redefining the canonical commutation relations:
\eqn\cutoff{
[\alpha_{nj},\alpha_{mk}]=
n(1-\epsilon)^{|n|} \delta_{j,k}\delta_{n+m,0}.
}
Note that our redefinition of the canonical commutation
relations will cause a modification of \lrid{}.  To avoid
this, we make a compensating redefinition of the oscillator
part of the Dirichlet state:  $|D^{\prime}\>_{oscillator}
\rightarrow e^{\sum_{n=1}^{\infty} {1\over n} (1-\epsilon)^n
  \balpha^{\dagger}_n\cdot \alpha^{\dagger}_n}
  |0\>$.

Our typical term is now
\eqna\typiv
$$\eqalignno{
|typ\> =&
    \epsilon^{\nu^2_1 + \cdots \nu^2_n}
    z^{\nu^2_1}_1 \cdots z^{\nu^2_n}_n \cr
  &  :e^{-i\nu_1\cdot M^t \cdot X(z_1)}: \cdots
        : e^{-i\nu_n\cdot M^t \cdot X(z_n)}:\;\;
        : e^{-i\nu_n\cdot X(z_n)}: \cdots
        : e^{-i\nu_1\cdot X(z_1)}:
|\pD\>.
}
$$
The next step is to shuffle the exponentials so that
exponentials which depend on the same coordinate are next
to each other.  By careful application of \CampHaus{},
one may show that
\eqn\shuffle{
:e^{i\nu_\cdot X(z)}: :e^{i\mu\cdot X(w)}: =
:e^{i\mu\cdot X(w)}: :e^{i\nu\cdot X(z)}: \;\;
e^{i\pi \nu\cdot\mu \;\sgn_{\epsilon}{(\sigma_z - \sigma_w)}},
}
where $z=e^{2\pi i \sigma_z /T}$,
$w=e^{2\pi i \sigma_w /T}$, and $0 \leq \sigma_z,\sigma_w \leq T$.
The function $\sgn_{\epsilon}(x)$ is a smoothed version
of the sign function $\sgn(x)= {x\over |x|}$;  the transition from
$\sgn_{\epsilon}(x) = -1$ to $\sgn_{\epsilon}(x)=1$ occurs
mostly over the interval $-\epsilon < x < \epsilon$.  As
$\epsilon\rightarrow 0$, $\sgn_{\epsilon}{(x)} \rightarrow \sgn{(x)}$.

We would also like to combine exponentials which depend on
the same coordinate.  With the aid of \CampHaus{}, we see that
\eqn\fuse{
:e^{i\nu_\cdot X(z)}: :e^{i\mu\cdot X(z)}: =
(\epsilon z)^{\nu\cdot\mu}
:e^{i(\nu+\mu)\cdot X(z)}:.
}
Applying \shuffle{} and \fuse{}, our typical term becomes
\eqna\typv
$$\eqalignno{
|typ\> =\;\;&
  \epsilon^{{1\over 2} \sum_k \phi^2_k}\;\;
  z^{{1\over 2} \phi^2_1} _ 1 \cdots
  z^{{1\over 2} \phi^2_n} _ n \cr
& \exp{\Bigl( {i\pi\over 2} \sum_{j>k} \phi_j \cdot (1-B) \cdot
       \phi_k \;\sgn_{\epsilon}{(\sigma_j-\sigma_k)} \Bigr) } \cr
& :e^{i\phi_1\cdot X(z_1)}: \cdots
   :e^{i\phi_n\cdot X(z_n)}:
|\pD\>, &\typv{}\cr
}$$
where we have introduced the convenient quantity
$\phi_k = -(1+M)\cdot\nu_k = -2(1-B)^{-1}\cdot\nu_k$.
Armed with equation \typv{}, we can now address the question
of conformal invariance.  Requiring conformal invariance will
lead us to a natural Kac-Moody algebra underlying our problem:
the exponentials of $X$ will be identified as Kac-Moody currents,
and the vectors $\phi_j$ will be the root vectors.
Once the algebraic structure is fleshed out, we will return
to expression \pertii{} and evaluate $|V\>$ explicitly in
terms of Lie algebraic quantities.

\subsec{Requirements for Conformal Invariance}
We will now impose the following two constraints:
\eqna\confconst
$$\eqalignno{
\phi^2_k = 2 &&\confconst{a}\cr
 {1\over 2}\phi_j\cdot (1-B) \cdot\phi_k \in\BZ. &&\confconst{b}\cr
}$$
With some work, we will see that \confconst{a} and \confconst{b}
lead to conformal invariance of $|V\>$.  We do not claim
that these are necessary conditions for conformal invariance;
indeed, calculations in \ckmy{} identify conformal theories
that do not obey our constraints.
Nevertheless, we choose to impose \confconst{a} and \confconst{b}
because they lead to a large class of conformal theories
where we can compute the boundary state exactly.

Equation \confconst{a} ensures that the exponentials in \typv{}
are weight one operators.  It also ensures that \typv{} will
have an overall factor of $\epsilon^n$.  Looking back at
equation \pertii{}, we see that we will have precisely one
factor of $\epsilon$ for each factor of $V_{\omega}$.
We can now dispose of $\epsilon$:  we rescale the bare couplings
by ${1\over\epsilon}$ and send $\epsilon \rightarrow 0$.

Equation \confconst{b} allows us to
replace the $\sgn{}$ function in \typv{} with a factor of unity.
Evidence from the $N=2$ version
of this problem \ckmy{} suggests that conformal
invariance holds when the $\sgn{}$ functions drop out.
Setting $\sgn{}=1$ in \typv{} also
disentangles the integrals over the $z$ coordinates, allowing us
to compute $|V\>$ very explicitly.

Noting that \confconst{b} holds for all $j$ and $k$, we can take the
transpose and obtain
\eqn\conftrans{
{1\over 2}\phi_j\cdot (1+B) \cdot\phi_k \in\BZ,
}
which can be combined with \confconst{b} to yield
\eqn\intprod{
\phi_j\cdot\phi_k \in\BZ.
}
Combined with \confconst{a}, \intprod{} tells us that
$\phi_j\cdot\phi_k\in \{0,\pm 1,\pm 2\}$.

These strong restrictions on $\phi_j$ immediately bring to mind the
root systems for Lie algebras.  The root vectors of any simply laced,
semi-simple Lie algebra obey exactly the same constraints as the
$\phi_j$.  Conversely, given a set of vectors $\phi_j$ subject
to the constraints \confconst{a} and \intprod{}, it can be proven
that the set $\{\phi_j\}$ is a subset of the set of root vectors
of some simply laced, semi-simple Lie algebra.  These facts may be
found in standard texts such as \refs{\cornwell,\var}.

Let us label the standard objects that arise for our Lie algebra:
call the algebra $g$, let $\psi_j$ denote the simple roots,
and let $\cR$ denote the set of roots.
The root lattice $\Lambda_{\cR}$ corresponds to the matrix consisting
of the $\psi_j$ as column vectors, and the weight lattice
$\Lambda_{\cW}$ is simply the dual of the root lattice.
The Cartan matrix $A$ is given by $A=\Lambda^t_{\cR}\Lambda_{\cR}$.
Note that we use the same symbols for a lattice and the matrix
of basis vectors for the lattice;  for a summary of lattice conventions
and facts, see the appendix.

So far, we have found that $\phi_j \in \cR$, but we still have
equation \confconst{b} to deal with.  It is convenient to recast
\confconst{b} in the form
\eqn\Blatrel{
{1\over 2}(1-B) \Lambda_{\cR} \subset \Lambda_{\cW}.
}
In other words,
we must have
${1\over 2} \Lambda^{-1}_{\cW} (1-B)\Lambda_{\cR} = D$,
where $D$ is a matrix with integer entries.  Solving for $B$, we find
\eqn\Bsolve{
B = 1-2\Lambda_{\cW}D\Lambda^t_{\cW}.
}
Requiring $B$ to be antisymmetric tells us that $D$ must satisfy
the equation $D+D^t=A$.  Noting that the Cartan matrix $A$ is
symmetric and has $2$ in all the diagonal entries, we can solve
for $D$:
\eqn\Dsolve{
D=1+\up{(A)} + F,
}
where $F$ is an arbitrary anti-symmetric matrix with integer entries,
and $\up{(A)}$ is the upper half of the matrix $A$ (i.e., take $A$
and set all entries on the diagonal and below to zero).

We now have the complete solution of our conformal constraints
\confconst{a} and \confconst{b}.  First, we pick any simply laced,
semi-simple Lie algebra $g$.  The $\phi_j$ are root vectors of $g$,
and the possible values for $B$ are enumerated by \Dsolve{}
and \Bsolve{}.  The original quantities $\omega_k$ which appear
in our action are given by a linear transformation of the root
vectors:  $\{\omega_k\} = {1\over 2}(1-B)\cR \subset \Lambda_{\cW}$.
It is interesting
to note that the gauge field is indispensable for realizing
all algebras other than direct sums of $su(2)$:  if we set $B=0$,
equation \Bsolve{} tell us that $D = {1\over 2}\Lambda^t_{\cR}
\Lambda_{\cR} = {1\over 2}A$.  Since the Cartan matrix $A$ has
entries of $-1$ for all algebras other than $su(2)$, $D$ will not
be the required integer matrix unless $g$ is a direct sum of $su(2)$
components.

\subsec{Compactness Constraints}
Let us consider the constraints due to compactification of $X$,
namely equations \Bsense{} and \omsense{}.  Recalling the definition
$\phi_k = -(1+M)\cdot\nu_k$, we can rewrite \omsense{} in terms of
the root lattice of $g$:
\eqn\Vcompact{
{1\over\sqrt{2}}(1-B)\Lambda_{\cR} \subset \Gamma^*.
}
For convenience, we also redisplay \Bsense{}:
\eqn\Bcompact{
{1\over 2} B \Gamma \subset \Gamma^{*}.
}

We need to address the following question:  given our choice of
Lie algebra $g$ and our solution for $B$, what are the possible
compactification lattices allowed by \Vcompact{} and \Bcompact{}?
Clearly we can always choose $X$ to be non-compact, regardless of
$B$, because $\Gamma^*=\BR^N$ in that case.

For compact $X$, equation \Vcompact{} implies that
\eqn\Gdef{
{1\over\sqrt{2}}\Gamma^t (1-B) \Lambda_{\cR} = G,
}
where $G$ is a matrix with integer entries.  Using \Bsolve{}
to eliminate $B$, we can solve for $\Gamma$:
\eqn\Gamsolve{
\Gamma = {1\over\sqrt{2}} \Lambda_{\cR} (D^{-1})^t G^t.
}

Equation \Bcompact{} implies that ${1\over 2}\Gamma^t B \Gamma$
is a matrix with integer entries.  Using \Bsolve{} and \Gamsolve{}
to eliminate $B$ and $\Gamma$, we get the condition
\eqn\Gconst{
{1\over 4}G\left(D^{-1}-(D^{-1})^t\right) G^t = \hbox{integer matrix}.
}
Any integer matrix $G$ which satisfies \Gconst{} can be plugged
into \Gamsolve{} to obtain a compactification lattice consistent
with our constraints \Vcompact{} and \Bcompact{}.  Clearly there
are many possible choices for $G$;  for example, we can always
use $G=2\sqrt{k}D$ or $G=2\sqrt{k}D^t$ with $k\in\BZ$.

\subsec{$|\pD\>$ in terms of Kac-Moody representations}
When we defined $|\pD\>$ in equation \Dprimedef{},
we did not make a definite choice for $\Upsilon$, the lattice
of momenta.  Our definition of $|\pD\>$ requires that
${1\over\sqrt{2}}(1+B)\Gamma\subset\Upsilon$, and the conversion
of right-movers into left-movers in equation \lrid{} works on
the assumption that $\omega_k\in\Upsilon$.  We can satisfy these
requirements by choosing $\Upsilon = \Lambda_{\cW}$.  To see
this, first take the dual of equation \Vcompact{}:
\eqn\Vcompactd{
{1\over\sqrt{2}}(1+B)\Gamma \subset \Lambda_{\cW}.
}
Next, recall that $\omega_k\in  {1\over 2}(1-B)\cR\subset
{1\over 2}(1-B)\Lambda_{\cR}$.  Looking at equation \Blatrel{},
we see that $\omega_k \in \Lambda_{\cW}$.

Once we set $\Upsilon=\Lambda_{\cW}$, we can reinterpret
$|\pD\>$ in terms of Kac-Moody representations using
the well-known vertex operator construction \go{}.  The level 1
Kac-Moody algebra $\hat{g}$ (i.e., the one corresponding to the
Lie algebra $g$) can be written in terms of $\partial X(z)$
and exponentials
of $X(z)$.  The vertex operator construction requires us to
restrict $p\in\Lambda_{\cW}$, which is exactly the case with
$|\pD\>$.  Each highest weight representation of $\hat{g}$
appears exactly once in the chiral Hilbert space of $X(z)$.
These facts allow us to rewrite $|\pD\>$ in the convenient
form
\eqn\DprimeKM{
|\pD\> = 2^{-N/4} \sqrt{\det(1+B)\vol{(\Gamma)}}
  \sum_{\mu\in\cW \atop s} |\mu , s \> \overline{|\mu , s \>},
}
where $\mu$ runs over the highest weights $\cW$ of the level 1
highest weight representations of $\hg$, and $s$ labels the states
of the representation $\mu$.

\subsec{Vertex Operators and Cocycles}
We will now use the vertex operator construction \go{} of the level
1 Kac-Moody algebra $\hg$ to turn the exponentials in \typv{} into
Kac-Moody currents.  The algebra $\hg$ has the commutation relations
\eqna\gcom
$$\eqalignno{
[H_{ni},H_{mj}] &= n\;\delta_{ij}\delta_{n+m,0}, \cr
[H_n,E^{\phi}_m] &= \phi E^{\phi}_{n+m},\cr
[E^{\phi}_n,E^{\pphi}_m] &=
  \cases{  \epsilon(\phi,\pphi) E^{\phi+\pphi}_{n+m},
               &if $\phi+\pphi$ is a root;\cr
           \phi\cdot H_{n+m} + n\delta_{n+m,0},
               & if $\phi + \pphi=0$; \cr
           0, &otherwise.\cr
   }
}$$
$H_n$ are the Cartan operators, $E^{\phi}_n$ are the ladder operators,
$\phi$ and $\pphi$ are root vectors, and $\epsilon(\phi,\pphi)$ takes
on the values $\pm 1$.  The Cartan operators are simply
the modes of $\partial X(z)$:  $H_{ni} = \alpha_{ni}$, and $H_{0i} = p_i$.
In terms of currents, we have $H(z) = \sum_{n=-\infty}^{\infty}
H_n z^{-n-1} = i\partial X(z)$.  The ladder operators are constructed
from the modes of vertex operators:
\eqn\vertexdef{
E^{\phi}(z) = \sum_{n=-\infty}^{\infty} E^{\phi}_n z^{-n-1} =
\; :e^{i\phi\cdot X(z)}: \; c_{\phi}(p).
}
The chiral cocycle factor $c_{\phi}$ depends purely on the momentum operator
$p$, which we explicitly indicate.  Without the cocycles, the commutation
relations for $\hg$ do not work out properly.  Following the treatment
in \go{} , we can explicitly construct the cocycles in the form
\eqn\codef{
c_{\phi} (p) =
\sum_{\pphi\in\Lambda_{\cR}}\sum_{\mu\in \cW}
  \epsilon(\phi,\pphi) |p=\pphi + \mu\> \<p=\pphi+\mu|.
}
Note that the sum over momenta is split into a sum over highest weight
representations, labeled by $\mu$, and the states in those
representations, labeled by $\pphi$.  The coefficients $\epsilon(
\phi,\pphi)$ are the same ones that appear in the commutation
relations for $\hg$.  Since we are free to change the definition
of $E^{\phi}_n$ by a factor of $-1$, the function $\epsilon$ is
not entirely fixed.  Any function obeying the following constraints
will do:
\eqna\epconst
$$\eqalignno{
\epsilon(\alpha,\beta) & \in \{1,-1\}, &\epconst{a}\cr
\epsilon(\alpha,\beta) & = (-1)^{\alpha\cdot\beta}\epsilon(\beta,\alpha),
  &\epconst{b}\cr
\epsilon(\alpha,\beta) \epsilon(\alpha+\beta,\gamma) &=
  \epsilon(\alpha,\beta+\gamma)\epsilon(\beta,\gamma),
   &\epconst{c}\cr
\epsilon(\alpha,0) = \epsilon(0,\alpha) & =1,
   &\epconst{d}\cr
\epsilon(\alpha,-\alpha) &=1,& \epconst{e}\cr
}$$
for arbitrary $\alpha,\beta,\gamma \in \Lambda_{\cR}$.

A good first guess is
\eqn\eptwiddef{
\tepsilon (\phi,\pphi) = e^{i{\pi\over 2} \phi\cdot (1+B)\cdot\pphi}.
}
It is easily verified that $\tepsilon$ satisfies all the conditions
except for \epconst{e}.  This may be remedied by defining
\eqn\etadef{
\eta_{\phi} \in \{1,-1\} \quad\hbox{chosen so that}
  \quad \eta_{\phi}\eta_{-\phi}=\tepsilon(\phi,-\phi),
}
and setting
\eqn\epdef{
\epsilon(\phi,\pphi) = \eta_{\phi} \; \eta_{\pphi} \;
 \eta_{\phi+\pphi} \;
  \tepsilon(\phi,\pphi).
}

For later convenience, we would like to fix the value of $\eta_{\phi}$
for the cases when $\phi$ is a root vector.  Noting that
$\eta_{\phi}\eta_{-\phi}=\tepsilon(\phi,-\phi)= -1$
 for all root vectors $\phi$, we take
\eqn\defetaroot{
\eta_{\phi} = \cases{1,& if $\phi$ is a positive root;\cr
        -1,& if $\phi$ is a negative root.\cr
      }
}

Finally, we can use the fact that $c_{\phi}(p)^2=1$ to re-express
\vertexdef{}:
\eqn\solvevertex{
:e^{i\phi\cdot X(z)}: \; = E^{\phi}(z) c_{\phi}(p).
}

\subsec{Final Result for the Boundary State $|V\>$}
Applying the results of the previous subsections, we can now
write the typical term \typv{} in terms of Kac-Moody currents:
\eqna\typvi
$$\eqalignno{
|typ\> =&  z_1 \cdots z_n \;\;
  e^{i{\pi\over 2} \sum_{j>k} \phi_j\cdot (1-B) \cdot \phi_k} \cr
& E^{\phi_1}(z_1) c_{\phi_1}(p) \cdots
  E^{\phi_n}(z_n) c_{\phi_n}(p) |\pD\>.
}$$
Recall that we have eliminated $\epsilon$ by rescaling the
bare couplings $V_{\omega}$.

The ladder operator $E^{\phi}(z)$ carries momentum $\phi$, so
we can move the cocycles past the ladder operators, providing
we shift the momenta:
\eqna\typvii
$$\eqalignno{
|typ\> =& z_1 \cdots z_n \;\;
  e^{i{\pi\over 2} \sum_{j>k} \phi_j\cdot (1-B) \cdot \phi_k} \cr
& E^{\phi_1}(z_1) \cdots E^{\phi_n}(z_n)
  c_{\phi_1}(p+\phi_2+\cdots + \phi_n)
  c_{\phi_2}(p+\phi_3+\cdots + \phi_n) \cdots
  c_{\phi_n}(p) |\pD\>.
}$$
Shifting $p$ simply means that
\eqn\pshift{
c_{\phi}(p+\beta) = e^{-iq\cdot\beta} c_{\phi}(p)
  e^{iq\cdot\beta} =
  \sum_{\pphi\in\Lambda_{\cR}}\sum_{\mu\in \cW}
  \epsilon(\phi,\pphi+\beta) |p=\pphi + \mu\> \<p=\pphi+\mu|.
}
Applying \eptwiddef{}, \etadef{}, and \epdef{}, we can simplify
the cocycle product, obtaining
\eqn\typviii{
|typ\> = z_1 \cdots z_n \;\;
 E^{\phi_1}(z_1) \eta_{\phi_1} \cdots
 E^{\phi_n}(z_n) \eta_{\phi_n}\;
U_{\sum_k \phi_k} (p) |\pD\>.
}
All that remains of the cocycles are the numbers $\eta_{\phi}$
and the operator
\eqn\Udef{
U_{\phi}(p) =
\sum_{\pphi\in\Lambda_{\cR}}\sum_{\mu\in \cW}
  \eta_{\pphi}\eta_{\phi+\pphi}
  e^{i{\pi\over 2} \phi\cdot (1+B) \cdot \pphi}
 |p=\pphi + \mu\> \<p=\pphi+\mu|.
}

Since $p|\pD\> = \bp|\pD\>$, we can make the replacement
$U_{\phi}(p)|\pD\> = U_{\phi}(\bp)|\pD\>$, allowing us to
commute $U$ to the left of the ladder operators:
\eqn\typix{
|typ\> = \epsilon^n \; z_1 \cdots z_n \;\;
 U_{p-\bp}(\bp) \;
 E^{\phi_1}(z_1) \eta_{\phi_1}
 \cdots E^{\phi_n}(z_n) \eta_{\phi_n} |\pD\>.
}
Plugging $|typ\>$ back into the series expansion \pertii{},
we see that we can perform the integrals over $z_1,\cdots z_n$ and
resum the series into an exponential:
\eqna\Vresult
$$\eqalignno{
|V\> =&
P_{\sqrt{2}\Gamma}(p-\bp)S_BR
U_{p-\bp}(\bp) \cr
 & \exp{\left( -T \sum_{\phi\in\cR^+} \left(
  V_{-{1\over 2}(1-B)\phi}\eta_{\phi}
  E^{\phi}_0 +
  \bV_{-{1\over 2}(1-B)\phi} \eta_{-\phi} E^{-\phi}_0 \right) \right)}
|\pD\>.&\Vresult{}\cr
}$$

Performing the contour integrals has left us with just the
zero modes $E^{\phi}_0$, where
$\phi$ runs over the set of positive roots $\cR^+$.
Since the Kac-Moody fields are weight one, we have
\eqn\LKMcom{
[L_n,E^{\phi}_m] = -m E^{\phi}_{n+m}.
}
In particular, $[L_n,E^{\phi}_0] = 0$.  Since we already
know that $[L_n,S_B] = [L_n,R] = [L_n,p] =0$, it follows
that $(L_n-\bL _n)|V\> =0$.  As promised, $|V\>$ is conformally
invariant, regardless of the value of the couplings $V_{\omega}$.

The only renormalization needed in our problem is a trivial
rescaling of $V_{\omega}$ by the cutoff $\epsilon$, which
appeared when we normal ordered the exponentials in
\normord{}.  Let us define the renormalized potential
strengths in terms of the original (unrescaled) bare
couplings:
$\tV_{\phi} = i\epsilon T V_{-{1\over 2}(1-B)\phi}$
and
$\tV_{-\phi} = -i\epsilon T \bV_{-{1\over 2}(1-B)\phi}$
for positive roots $\phi$.
Our final result is
\eqn\Vfinal{
|V\> =
P_{\sqrt{2}\Gamma}(p-\bp)S_BR
U_{p-\bp}(\bp) \;
\exp{\left( i\sum_{\phi\in\cR} \tV_{\phi}
  E^{\phi}_0  \right)}
|\pD\>.
}
This expression for $|V\>$ has a very appealing simplicity:
we act on the basic state $|\pD\>$ with a series of unitary
operators, then project onto the space of allowed momenta.
Each unitary operator has an obvious interpretation:
$\exp{(i\sum\tV_{\phi} E^{\phi}_0)}$ is the Lie group element
corresponding to the potential $V(X)$.
$S_B$ is the $SO(N)$ rotation due
to the gauge field.
The operator $U$, which
only takes on the values $\pm 1$, serves to keep track
of what choice of signs we made in defining our Kac-Moody
operators.
Lastly, the reflection operator $R$ allows us to switch between
the Neumann boundary state and the more convenient Dirichlet
boundary state.

\newsec{Partition Functions}
Now that we have obtained the boundary state created by the
potential $V(X)$, we can compute the cylinder partition
function with various boundary conditions on the opposite
end.  Due to the algebraic structure of the boundary state,
all results can be expressed as traces over the Kac-Moody
representation space.  We will use the notation
$\Tr_{\hg}(\cdots)$, meaning that we trace over all the states
of each highest-weight representation of $\hg$.

\subsec{Neumann Case}
Putting $|N\>$ at one end and $|V\>$ at the other,
the partition function is
\eqn\ZNVi{
Z_{NV} = \<N|e^{-{2\pi\over T} l (L_0 + \bL_0)} |V\>.
}
Setting $w=e^{-2\pi l / T}$ and applying \Vfinal{}, we get
\eqna\ZNVii
$$\eqalignno{
Z_{NV} &= \<N|w^{2L_0}
  P_{\sqrt{2}\Gamma}(p-\bp)S_BR
U_{p-\bp}(\bp) \;
e^{ i\sum \tV_{\phi}
  E^{\phi}_0  }
|\pD\>  \cr
&= \<N|w^{2L_0} R P_{ {1\over\sqrt{2}}\Gamma}(p)
  U_{(M-1)\cdot p}(p) S_B e^{ i\sum \tV_{\phi} E^{\phi}_0  }
  |\pD\> \cr
&= 2^{-N/2} \sqrt{\det{(1+B)}} \vol{(\Gamma)}
  \Tr_{\hg} \left( w^{2L_0} P_{ {1\over\sqrt{2}}\Gamma}(p)
     U_{(M-1)\cdot p}(p) S_B e^{ i\sum \tV_{\phi} E^{\phi}_0  }
  \right). \cr
}$$

In the special case where $g=su(2)$ and $B=0$, our result simplifies
to
\eqn\ZNViii{
Z_{NV} = 2^{-1/2} \vol{(\Gamma)} \Tr_{\hg} \left(
   w^{2L_0} P_{ {1\over\sqrt{2}}\Gamma}(p)
   e^{ i\sum \tV_{\phi} E^{\phi}_0  }\right).
}
If we specialize further by taking the compactification lattice
to be $\Gamma = {1\over\sqrt{2}} \Lambda_{\cR} = \BZ$, we get
a sum over the characters of $\hg$:
\eqn\ZNViv{
Z_{NV} = 2^{-1/2}\Tr_{\hg} \left(w^{2L_0}
    e^{ i\sum \tV_{\phi} E^{\phi}_0  }\right).
}

\subsec{Dirichlet Case}
Putting $|D\>$ at one end and $|V\>$ at the other,
the partition function is
\eqna\ZDVi
$$\eqalignno{
Z_{DV} &=
  \<D|w^{L_0+\bL_0}|V\> \cr
&=
  \<D|w^{2L_0}P_{\sqrt{2}\Gamma}(p-\bp)S_BR U_{p-\bp}(\bp)\;
  e^{ i\sum \tV_{\phi}E^{\phi}_0  }|\pD\>  \cr
 &=
  2^{-N/2} \sqrt{\det{(1+B)}} \Tr_{\hg}
  \left( w^{2L_0} P_{{1\over\sqrt{2}}\Gamma^*}(p)
  U_{-(1+M)p}(p) S_B R
  e^{ i\sum \tV_{\phi}E^{\phi}_0  } \right).
}$$

\subsec{Gauge Field and Potential at Both Ends}
For our final example, we will put the same
gauge field state $|B\>$ and Lie group
$g$ at both ends, but we will allow the values of the
$V_{\omega}$ to be different.  Physically speaking, we are
studying an open string whose ends feel periodic potentials
$V^L(X)$ and $V^R(X)$ and a single uniform field $B$.
It is important to note that the ends are {\it oppositely}
charged as
far as the gauge field is concerned.
The partition function is
\eqna\ZVVi
$$\eqalignno{
Z_{VV} &=
  \<V^L|e^{-{2\pi\over T} l (L_0 + \bL_0)}|V^R\> \cr
&=
 \<\pD|e^{ -i\sum \tV^L_{\phi}E^{\phi}_0  }
  U^{\dagger}_{p-\bp}(\bp)
  R^{\dagger}S^{\dagger}_B
  P_{\sqrt{2}\Gamma}(p-\bp)
   w^{2L_0}
  P_{\sqrt{2}\Gamma}(p-\bp)S_B R U_{p-\bp}(\bp) \;
  e^{ i\sum \tV^R_{\phi}E^{\phi}_0  }|\pD\>,  \cr
}$$
again with $w=e^{-{2\pi\over T} l}$.
The rotations $S_B$ and $S^{\dagger}_B$
for the two ends are conjugate, indicating that the string
does indeed have zero net charge with respect to the
gauge field.   After all the conjugate operators cancel
out, we get
\eqna\ZVVii
$$\eqalignno{
Z_{VV} &=
  \<\pD| w^{2L_0} e^{ -i\sum \tV^L_{\phi}E^{\phi}_0  } \;\;
   P_{\sqrt{2}\Gamma} (M^t p +\bp) \;
   e^{ i\sum \tV^R_{\phi}E^{\phi}_0  }|\pD\> \cr
&=
  2^{-N/2}\det{(1+B)}\vol{(\Gamma)} \cr
&  \sum_{\mu\in\cW\atop s} \<\mu ,s|
  w^{2L_0}
 e^{ -i\sum \tV^L_{\phi}E^{\phi}_0  } \;
  P_{\sqrt{2}\Gamma} (M^t p + p(\mu,s))
  \;e^{ i\sum \tV^R_{\phi}E^{\phi}_0  }
  |\mu,s\>, &\ZVVii{}\cr
}$$
where $p(u,s)$ is defined to be the momentum of the
state $|\mu,s\>$, i.e., $p|\mu,s\> = p(\mu,s)|\mu,s\>$.
The presence of $p(\mu,s)$ inside the projection
operator prevents us from writing \ZVVii{} as
a trace.  In the special case where $\tV^{L}_{\phi}=0$,
we can replace $p(\mu,s)$ with $p$, giving us
\eqn\ZBV{
Z_{BV} =
  2^{-N/2}\det{(1+B)}\vol{(\Gamma)}
  \Tr_{\hg} \left( w^{2L_0}
    P_{{1\over\sqrt{2}}(1+B)\Gamma}(p)
   \;e^{ i\sum \tV^R_{\phi}E^{\phi}_0  }
   \right).
}

\newsec{Correlation Functions and the S-Matrix}
The boundary state $|V\>$ provides a complete specification
of the manner in which left-movers scatter into right-movers
when they hit the boundary.  The simplest way to extract
the scattering data is to compute the correlation functions
of $\partial X$ and $\bpar\bX$.  Let us define the
general correlation function on the cylinder
\eqn\Fdef{
F(z,\brho) = \<\partial X_{i_1}(z_1) \cdots
   \partial X_{i_n}(z_n) \bpar \bX_{j_1}(\brho_1)
   \cdots \bpar \bX_{j_m}(\brho_m) \>.
}

In the boundary-state language, we have
\eqn\Fbound{
F(z,\brho) = {1\over Z_{CV}}
  \<C| w^{2L_0}
   \partial X_{i_1}(z_1) \cdots
   \partial X_{i_n}(z_n) \bpar \bX_{j_1}(\brho_1)
   \cdots \bpar \bX_{j_m}(\brho_m) |V\>,
}
where $w=e^{-2\pi l/T}$ and
$<C|$ represents some arbitrary boundary condition
at the other end of the cylinder.

Now recall equation \Vfinal{} which shows how we can
express $|V\>$ in terms of a projection operator
and unitary operators acting on $|\pD\>$.
The projection operator commutes with $\partial X$
and $\bpar \bX$, so we may move it to the left.
The unitary operators may also be moved to the left,
conjugating the $\partial X$ operators as they go by.
This leaves the $\bpar \bX$ operators next to the $|\pD\>$
state, which can be used to convert them into $\partial X$
operators.  More precisely, the relation is $\bX(\bz) |\pD\>=
-X(1/\bz)|\pD\>$, so we have
\eqna\Fboundii
$$\eqalignno{
F(z,\brho) =& {1\over \brho^2_1\cdots \brho^2_m Z_{CV}}
  \<C| w^{2L_0}
P_{\sqrt{2}\Gamma}(p-\bp)S_BR
U_{p-\bp}(\bp) \;
e^{ i\sum \tV_{\phi}
  E^{\phi}_0  } \cr
&
 \partial \tX_{i_1}(z_1)\cdots
  \partial \tX_{i_n}(z_n)
 \partial X_{j_m}(1/\brho_m) \cdots
 \partial X_{j_1}(1/\brho_1) |\pD\>. &\Fboundii{}
}$$
The new operators $\tX$ are simply conjugated left-movers:
\eqn\deftX{
\partial \tX_i(z) = e^{ -i\sum \tV_{\phi} E^{\phi}_0  }
    R S^{\dagger}_B \;\partial X_i(z)\;
    S_B R e^{ i\sum \tV_{\phi} E^{\phi}_0  }.
}
Conjugation by $S_B$ results in an $SO(N)$ rotation of
the $\partial X_i$.  Conjugation by $R$ simply introduces
a minus sign.
Lastly, conjugation by the group element
$e^{ i\sum \tV_{\phi} E^{\phi}_0 }$ results in a linear
combination of the Cartan operators $\partial X_i$
and ladder operators $E^{\phi}$.  Introducing the
coefficients $b^V_{jk}$ and $c^V_{j\phi}$ to describe
the group rotation due to $V(X)$, we have:
\eqn\deftXii{
\partial \tX_i(z) =
  -M^t_{ij} \left( b^V_{jk} \partial X_k (z) +
     c^V_{j\phi} E^{\phi} (z) \right).
}

For the sake of simplicity, we will now expand the cylinder
into a half-plane by taking $T \rightarrow\infty$ and
$l \rightarrow\infty$.  Since $w \rightarrow 0$ as
$l\rightarrow\infty$, only the vacuum amplitude in
equation \Fboundii{} survives.  Furthermore, since
$z=e^{2\pi (\tau+ i\sigma)/T}$, $z\rightarrow 1$
and $\bz\rightarrow 1$ as $T\rightarrow\infty$.
It makes sense to change over to coordinates that
are more convenient for the half-plane geometry,
i.e., we switch to the coordinates $z=\tau + i\sigma$.
In fact, we will go one step further and switch
to the {\it open-string} picture on the half-plane:
we use coordinates $z=\sigma + i\tau$, and the boundary
is the imaginary axis $\sigma=0$.  The correlation functions are
given by
\eqna\solvecorr
$$\eqalignno{
\<0|\partial X_{i_1} (z_1) \cdots \partial X_{i_n} (z_n)
   \bpar \bX_{j_1}(\brho_1) \cdots \bpar \bX_{j_m} (\brho_m)
|0\>_{open} = \cr
\<0| \partial \tX_{i_1} (z_1) \cdots \partial \tX_{i_n} (z_n)
   \partial X_{j_m}(-\brho_m) \cdots
   \partial X_{j_1}(-\brho_1) |0\>_{free}.
   &&\solvecorr{} \cr
}$$
The left-hand side of equation \solvecorr{} is a correlation
function for open-string fields propagating on the half-plane
$\sigma > 0$ with the nontrivial action $S_{gauge}+S_{pot}$
acting at the spatial boundary $\sigma=0$.  The right hand
side of \solvecorr{} is the same correlation function expressed
entirely in terms of left-movers.  Since the boundary interaction
only affects correlations between different chiralities,
the right-hand side of \solvecorr{} can be considered a correlation
function for chiral free fields on the full plane.
Since $\tX$ is just a linear combination of $\partial X$
and vertex operators, one may easily evaluate the right hand
side of \solvecorr{} with coherent state methods.
Note that right-moving coordinates $\brho_k$ are reflected
to the image points $-\brho_k$;  singularities between
left-movers and right-movers only occur when the coordinates
coincide on the imaginary axis $\sigma=0$.

The content
of equation \solvecorr{} is conceptually simple:  we start
with both chiralities in the half-plane
(i.e., the
left hand side of \solvecorr{}), then we convert to a single chirality in
the whole plane by reflecting the right-movers into image
left-movers.  The effect of the boundary interaction is
summarized by the replacement $\partial X \rightarrow
\partial \tX$.  The operator
$e^{ -i\sum \tV_{\phi} E^{\phi}_0  }
    R S^{\dagger}_B $
is the boundary S-matrix operator;  the action of the
S-matrix on our basic operators $\partial X$ is
summarized by \deftXii{}.

\newsec{Unitarity}
Some recent papers \refs{\ali,\cklm} have commented on an apparent
unitary violation when particles scatter from a boundary
in conformal field theory.  Although the particular models
studied in \ali{} and \cklm{} are different, the general
features of the problem are the same:  the S-matrix
is explicitly computed, showing that certain ingoing
states (i.e., combinations of left-movers) have
less than unit probability of scattering into any
combination of outgoing (i.e., right-moving) states.
In some cases, the ingoing states seem to disappear
entirely!  The resolution of the paradox also takes
the same general form in both cases:  the Hilbert
space of ingoing and outgoing states must be enlarged
to include soliton states that were not originally
included.  In the enlarged Hilbert space, the S-matrix
is perfectly unitary.

The model considered in this paper is a generalization
of the $N=1$ case studied in \cklm{} , and so it is not
surprising to find that the same unitarity question
arises, with essentially the same solution.
Let us begin by working in the Hilbert space
built from the the operator $\partial X$.
The scattering relation \deftXii{} shows us
that outgoing states contain the ladder operators
$E^{\phi}$.  Since the ladder operators carry
momentum while the $\partial X$ operators don't,
it is clear that part of the outgoing state
will be orthogonal to our Hilbert space built
from $\partial X$.  The solution is obvious:
we need to study scattering in the Hilbert space
which includes the soliton states created by the
ladder operators.

One possible description of
our enlarged Hilbert space is in terms of mutually
orthogonal sectors:  each sector is labeled by
a momentum $\mu$ in the weight lattice $\Lambda_{\cW}$.
The sector with momentum $\mu$ is constructed
by letting the modes of $\partial X$ act on
the $\mu$-ground state $e^{iq\cdot\mu}|0\>$.
This
enlarged Hilbert space is nothing other than the direct sum
of the highest-weight representations of the algebra
$\hg$, with each representation appearing exactly
once.
The ladder operators $E^{\phi}$ carry momentum
equal to the root vector
$\phi$, so they create solitons interpolating between different
$\mu$-sectors.

Since the root lattice is a proper sublattice
of the weight lattice, we do not actually have to
use all the $\mu$-sectors to describe scattering.
The minimal choice is to use only the sectors
labeled by root lattice momenta, in which case
the Hilbert space is simply the highest-weight
representation of $\hg$ built from the vacuum state
$|0\>$.

\newsec{Conclusions}
In this paper we have identified and solved a family
of bosonic boundary conformal field theories with
integer central charge.  These theories exhibit a
natural Kac-Moody current algebra structure, permitting
us to compute the exact boundary state, partition
function, and correlation functions.
We find that conformal invariance is independent
of the strength of the boundary potential, allowing
us to wander over an entire manifold of fixed points
by varying the couplings $V_{\omega}$.
Our results extend the $su(2)$-based calculations
of \refs{\cklm,\ckmy} to the entire A-D-E series of simply
laced semi-simple Kac-Moody algebras.  It is worth
noting that the gauge field part of the boundary
interaction is absolutely vital for realizing
all algebras other than $su(2)$.

These theories have at least
two possible applications.
In the context of string theory, our calculations
identify a series of solutions of the classical
open string equations:  the spacetime fields which
couple to the open string endpoints
are a uniform abelian gauge field and a periodic potential
for the tachyon.

Another application is the
dissipative quantum mechanics of a charged particle moving
in $N$ dimensions, subject to a periodic potential
and a magnetic field.  Our results provide
a complete description of the
critical behaviour of such systems, although
we have restricted ourselves to a special class
of magnetic fields and potentials.  Perturbative
calculations \ckmy{} for $c=2$ indicate that the
Kac-Moody structure is destroyed at generic values
of the magnetic field, so we do not expect any
simple extension of our results to the case of
arbitrary magnetic fields and potentials.

Finally, we offer the speculation that the type of
models considered in our paper may have applications
in edge current tunneling in the quantum Hall effect.
Our reasoning is the same as that presented in \ckmy{},
where a $c=2$ version of our model was studied.
In order to make any connection with the Hall effect,
it would appear that we need to find an integrable
deformation of our model.  We hope to address the
possibility of such an integrable model in future
work.

\appendix{}{Lattices}
A lattice on $\BR^N$ is a set of vectors of the form
\eqn\deflat{
\Lambda = \Bigl\{ \sum_{i=1}^{N} n_i\lambda_i
   \Bigm| n_i \in\BZ \Bigr\} ,
}
where the independent vectors $\lambda_1,\cdots \lambda_N$
are a basis for $\Lambda$.  It is useful to define a matrix
built out of the basis column vectors $\lambda_i$:
\eqn\defmat{
\Lambda_{matrix} = \bigl(\lambda_1 \lambda_2 \cdots \lambda_N
  \bigr).
}
 From now on, we will drop the `matrix' subscript and use the
single symbol $\Lambda$ for both the lattice and the matrix;
the context will make it clear which one is meant.

The unit cell of $\Lambda$ is the set
$\bigl\{\sum_{i=1}^{N}x_i\lambda_i
\bigm| 0\leq x_i < 1 \bigr\}$.
The lattice volume is defined to be the volume of the unit
cell, given by
\eqn\defvol{
\vol{(\Lambda)} = \sqrt{|\det{\left(\Lambda^t \Lambda\right)}|} =
  |\det{\Lambda}|.
}

The dual $\Lambda^*$ of the lattice $\Lambda$ is defined as
\eqn\defdual{
\Lambda^* = \Bigl\{v\in\BR^N \Bigm| v\cdot\lambda\in\BZ,
  \;\forall\lambda
  \in\Lambda \Bigr\},
}
Given a basis $\lambda_i$ for $\Lambda$, we can define a
canonical basis $\lambda^*_i$ for $\Lambda^*$ such that
$\lambda_i\cdot\lambda^*_j = \delta_{ij}$.  In terms of matrices,
we have the simple relation
\eqn\dualmat{
\Lambda^* = \left(\Lambda^{-1}\right)^t =
  \left(\Lambda^t\right)^{-1}.
}
The matrix representation makes it trivial to see that
$\vol{(\Lambda)} = \vol{(\Lambda^*)}^{-1}$.

The matrix representation also makes it simple to express
inclusion relations:
$$\displaylines{
v\in\Lambda  \;\Longleftrightarrow \;
  \Lambda^{-1}\cdot v \in \BZ^N \cr
\hbox{and}\cr
\Lambda\subset\Gamma \;\Longleftrightarrow \;
  \Lambda^* \supset \Gamma^* \;\Longleftrightarrow \;
  \Gamma^{-1} \Lambda \in \BZ^{N\times N}, \cr
}$$
where $\BZ^N$ is the set of vectors with $N$ integer components
and $\BZ^{N\times N}$ is the set of $N$ by $N$ matrices with
integer components.

As a notational convenience, it is
useful to broaden the definition of a lattice
to include any set of vectors closed under addition with
integer coefficients.  Under the broader definition, we can
think of the set $\Lambda =\{0\}$ as a degenerate
lattice.  This allows us to write down expressions for
compactified and uncompactified theories in the same
notation.

\newsec{Acknowledgements}

We would like to thank C. G Callan,
I. Klebanov, and J. Maldacena for useful discussions
and advice.  This work was supported in part by
DOE grant DE-FG02-91ER40671.

\listrefs
\end